\documentclass[12pt]{article}
\usepackage{amsfonts}
\usepackage{bm}
\usepackage{amsmath}
\usepackage{epsfig}

\usepackage{amssymb,lscape}
\usepackage[matrix,arrow,curve]{xy}

\newcommand{\bmat}{\left(\begin{array}}
\newcommand{\emat}{\end{array}\right)}

\def\gtrsim{\mathrel{\raise.3ex\hbox{$>$\kern-.75em\lower1ex\hbox{$\sim$}}}}

\def\-{\hphantom{-}}

\def\s2{\frac{1}{\sqrt2}}

\def\mg{m_{3/2}}
\def\mg2{m^2_{3/2}}

\def\Dsl{\,\raise.15ex\hbox{/}\mkern-13.5mu D} %this one can be subscripted

\def\be{\begin{equation}}
\def\ee{\end{equation}}
\def\bea{\begin{eqnarray}}
\def\eea{\end{eqnarray}}

\newcommand{\nn}{\nonumber}

\begin{document}

\pagestyle{plain}

%----------------------------------------------------------------------%
%  numbering equations with section number
%----------------------------------------------------------------------%
\makeatletter
\@addtoreset{equation}{section}
\makeatother
\renewcommand{\theequation}{\thesection.\arabic{equation}}
%----------------------------------------------------------------------%
%  title page
%----------------------------------------------------------------------%
\pagestyle{empty} \rightline{IPhT-t12/002}
\begin{center}
\LARGE{Gauged Double Field Theory \\[10mm]}
\large{Mariana Gra\~na and  Diego Marques
 \\[6mm]}
\small{{\it  Institut de Physique Th\'eorique, CEA/ Saclay}, \
91191 Gif-sur-Yvette Cedex, France.}
  \\[1cm]

\small{\bf Abstract} \\[0.5cm]\end{center}

{\small We find necessary and sufficient conditions for gauge invariance of the action of Double Field Theory (DFT) as well as closure of the algebra of gauge symmetries. The so-called weak and strong constraints are sufficient to satisfy them, but not necessary.  We then analyze compactifications of DFT on twisted double tori satisfying the consistency conditions. The effective theory is a Gauged DFT where the gaugings come from the duality twists. The action, bracket, global symmetries, gauge symmetries and their closure are computed by twisting their analogs in the higher dimensional DFT. The non-Abelian heterotic string and lower dimensional gauged supergravities are particular examples of Gauged DFT.

\newpage
%----------------------------------------------------------------------%
%  Resetting of counters
%----------------------------------------------------------------------%
\setcounter{page}{1}
\pagestyle{plain}
\renewcommand{\thefootnote}{\arabic{footnote}}
\setcounter{footnote}{0}
%----------------------------------------------------------------------%
%  Paper begins
%----------------------------------------------------------------------%

\tableofcontents

\section{Introduction}
Double Field Theory (DFT)  is a recent development that promotes a string  duality to a symmetry in field theory. It is manifestly invariant under global symmetries that include T-duality transformations. The original formulation was done in \cite{Hull:2009mi}, following previous ideas by Siegel \cite{Siegel:1993th} and Tseytlin \cite{Tseytlin} (see also \cite{Duff:1989tf}). Since then, DFT has been extended in many different ways \cite{Hohm}-\cite{Hohm:2011cp}. Some very recent related works on duality based constructions are \cite{WaldramR, Coimbra:2011nw}. Some reviews on the subject can be found in \cite{Zwiebach:2011rg}.

The theory is formally defined on a double space, that includes coordinates which are Fourier dual to momentum modes, plus T-dual coordinates  associated to winding. However, the current formulation of DFT is restricted by the so-called weak and strong constraints, that ensure gauge invariance and closure of the algebra. Combined, these constraints imply that the theory is only consistent on a slice of the double space parameterized by half of the coordinates, such that there always exists a frame in which locally the configurations do not depend on dual coordinates. Therefore, the doubling is only formal and the theory is not truly doubled. Since the constraints are covariant under the global symmetries, the theory can still be covariantly formulated.

Recently, flux compactifications of DFT were considered in \cite{Aldazabal:2011nj, Geissbuhler:2011mx}. It was realized that Scherk-Schwarz dimensional reductions \cite{Scherk:1979zr} of DFT lead to gauged supergravities in lower dimensions, in which the so-called non-geometric fluxes \cite{Shelton:2005cf} are purely geometric when analyzed from the point of view of the double space \cite{Hull:2004in, Dall'Agata:2007sr}. Interestingly, when the constraints are applied to the compactification ansatz, the gaugings in the effective theory are forced to satisfy relations that are stronger than the usual quadratic constraints (Jacobi identities) required for gauge consistency. The fact that the effective action is overconstrained, suggests that the constraints could be relaxed on the internal space.

In this paper, we explicitly show that it is possible to relax the weak and strong constraints. We first show that demanding gauge invariance and closure of the gauge transformations leads to a set of {\it invariance} and {\it closure} constraints. These constraints select subsets of fields and gauge parameters for which the gauge symmetries of DFT are consistent. Interestingly,  the weak and strong constraints are sufficient to satisfy the closure and invariance constraints,  but not necessary.

In \cite{Hohm:2011ex} it was shown that DFT can be deformed by gaugings that preserve the global covariance of the theory. The gaugings allow for non-Abelian gauge symmetries for vectors, and therefore this scenario is fruitful for embeddings of the Heterotic String \cite{Hohm:2011ex} and gauged supergravities \cite{Aldazabal:2011nj, Geissbuhler:2011mx} in DFT. Here we show that these deformed theories, that we call Gauged DFT (GDFT), can be obtained from Scherk-Schwarz flux compactifications of higher dimensional (ungauged) DFTs. In this sense, the gaugings are not introduced by hand, but arise from the duality twists of the compactification and therefore have a higher dimensional origin. We show that the action of GDFT, its global symmetries, bracket, gauge symmetries and their closure, and the constraints can all be obtained by twisting their analogs in the parent higher dimensional DFT. The twisted constraints are such that  their dependence on the gaugings always appears in the form of Jacobi-like quadratic constraints, plus terms that vanish if the strong and weak constraints are imposed on the  fields and gauge parameters of the effective lower dimensional theory. Since the quadratic constraints for the gaugings are less restrictive than the weak and strong constraints, this implies that they can be relaxed on the internal space (a similar situation occurs for the strong constraint in the Ramond-Ramond sector of (massive) Type II theories \cite{MassiveII}).

Finally, we show how the non-Abelian heterotic supergravity can be embedded in GDFT, by adding extra internal coordinates associated to the generators of the heterotic gauge symmetry, and twisting them to recover the structure constants of the non-Abelian algebra. Such a procedure is possible only after relaxing the constraints on the internal directions.

The paper is organized as follows. In Section \ref{ReviewDFT} we review the generalized metric formulation of DFT. We obtain the closure and invariance constraints required for consistency of gauge symmetries. Section \ref{GDFT} is devoted to analyze generalities of Scherk-Schwarz dimensional reductions of DFT.  In Section \ref{Examples} we review how to embed the $10$-dimensional non-Abelian heterotic String into GDFT. In Section \ref{conclusions} we summarize our main results and conclude.

\section{Double Field Theory}\label{ReviewDFT}

DFT is a field theory  invariant under a global symmetry group $G$ with a $K$-dimensional fundamental representation (with indices $M,N = 1,\dots, K$), and a symmetric metric $\eta_{MN}$. The coordinates $X^M$ form
fundamental vectors of $G$. To describe the NS degrees of freedom of string theory, one takes the group $G$ to be $O(K/2,K/2)$, but until Section \ref{Examples} we do not need to specify what $G$ is, we just need that it preserves a symmetric metric.

 The dynamical fields are the symmetric generalized metric
${\cal H}_{MN} \in G$, with inverse given by
$ {\cal H}^{MN} = \eta^{MP} {\cal H}_{PQ} \eta^{QN}$, and a   $G$ scalar $d$.
The generalized metric can be written in terms of a generalized $K$-bein ${\cal E}^a{}_M$
\be
{\cal H}_{MN} = {\cal E}^a{}_M S_{ab} {\cal E}^b{}_N \ ,
\ee
where $S_{ab}$ is the planar metric in $K$ dimensions (where $a,b = 1,\dots,K$) with appropriate signature (i.e., for the case $G = O(D,D+d)$, $S_{ab}={\rm diag}(-1,1...1;-1,1...1;1...1)$).

The $G$ transformations act as \be X'^M = U^M{}_P X^P\
, \ \ \ \  { {\cal E}'^a{}_M (X')} = (U^{-1})^{P}{}_{M}\  {\cal E}^a{}_P (X) \
, \ \ \ \   d'(X') = d (X) \
, \ \ \ \  U \in G  \ . \label{ODDninv} \ee

The dynamics of DFT is described by an action that can be written in a compact form (up to
total derivatives) in terms of a generalized Ricci scalar as \bea
S_{DFT}=\int d^{K}X\ e^{-2  d}\ {\cal R}
({ {\cal E}}, d)\ ,\label{action} \eea where ${\cal R}$ is defined
by \bea {\cal R}({ {\cal E}},  d)&=&4{ {\cal H}}^{  M N}\partial_{
M}\partial_{ N} d-\partial_{ M}\partial_{ N} {\cal H}^{  M  N}-4
{\cal H}^{  M N}\partial_{ M} d\partial_{  N} d +
 4\partial_{  M}  {\cal H}^{  M  N}\partial_{  N}   d \cr
&&+\ \frac18  {\cal H}^{  M  N}\partial_{  M} {\cal H}^{ K
L}\partial_{  N}  {\cal H}_{  K L}-\frac12 {\cal H}^{ M N}\partial_{
M} {\cal H}^{ K L}
\partial_{  K}  {\cal H}_{  N  L} \nn\\
&& + \ \frac{1}{2} \partial_M {\cal E}^a{}_P \partial^M {\cal E}^b{}_Q S_{ab}\eta^{PQ}\ .\label{R}
\eea
 Originally the coefficients in the first two lines  were chosen by hand in order to require invariance of the action under the symmetries, but it was shown that this action (with the last term included) is actually up to total derivatives the Ricci scalar corresponding to the generalized metric, or to be more precise to a torsion free  generalized connection \cite{Siegel:1993th,WaldramR,Hohm:2011cp}. The last term vanishes if we impose the strong constraint, given below in (\ref{strongconstraint}), and is not in the original generalized metric formulation of DFT but we include it to recover the results of \cite{WaldramR} and, as we will show in the next section, because it  provides the right contribution to the dimensionally reduced theory.

\subsection{Gauge symmetries and constraints}

The infinitesimal gauge transformations are generated by parameters $\xi^M(X)$ in the fundamental representation of $G$, and take the form
\bea \delta_{\xi}
d &=&   \xi^{  M}\partial_{  M}   d -
\frac 12 \partial_{  M}   \xi^{  M}  \ ,   \nn \\
\delta_{ \xi}  {\cal E}^a{}_M&=& \xi^P\partial_P
 {\cal E}^a{}_M
                  +\left(\partial_{  M} \xi^{  P}-\partial^{  P}
 \xi_{  M} \right)  {\cal E}^a{}_P
                 \ .\label{gaugeK} \eea
                 In general, one can define a gauge transformation for a generic tensor
 \be
 \delta_{ \xi}  {V}^M{}_N = \xi^P\partial_P
 V^M{}_N                  +\left(\partial^{  M} \xi_{  P}-\partial_{  P}
 \xi^{  M} \right)  V^P{}_N + \left(\partial_{ N} \xi^{  P}-\partial^{  P}
 \xi_ { N} \right)  V^M{}_P
                 \ .
 \ee
 Here we illustrate the transformations using a tensor with an upper and a lower index such that the generalization to an arbitrary tensor is easy to obtain.
 These transformations  define the so-called C-bracket
\be \left[ \xi_1, \xi_2\right]^{ M}_{\rm C} =
2 \xi_{[1}^{ N}
\partial_{  N}
  \xi^{  M}_{2]} -   \xi^{  P}_{[1} \partial^{  M}
  \xi_{2]  P} \ . \label{Cbracket} \ee
Demanding that the commutator of two transformations of an arbitrary given tensor $V^M{}_N$  behaves as a transformation itself
  \be \left[  {\delta}_{ \xi_1},  {\delta}_{ \xi_2} \right] V^{ M}{}_N =
  {\delta}_{[ \xi_1, \xi_2]_{\rm C}} V^{ M}{}_N  - F^M{}_N (\xi_1, \xi_2, V) \ ,\label{DefC}\ee
leads to the following first {\it closure} constraint\footnote{The index structure of the constraints (\ref{FirstIntermediate}) and (\ref{SecondIntermediate}) are those of the tensor acted upon  by the commutator  and Jacobiator.}
\be
F^M{}_N  (\xi_1, \xi_2, V) = \xi^Q_{[1} \partial^P \xi_{2]Q} \partial_P V^M{}_N + 2 \partial_P \xi^Q_{[1} \partial^P \xi_{2]N} V^M{}_Q + 2 \partial_P \xi_{[1 Q} \partial^P \xi_{2]}^M V^Q{}_N  = 0 \ .\label{FirstIntermediate}
\ee
When fields and gauge parameters are restricted in such a way that one can always find a frame in which the dependence on dual coordinates vanishes, this equation is automatically satisfied. However, different configurations could be considered for which this is not the case.

The C-bracket has a non-vanishing Jacobiator, which  can be written as
\be {\cal J}_{\rm C}^M (\xi_1, \xi_2, \xi_3) = 3[ \xi_{[1},[\xi_{2},\ \xi_{3]}]_{\rm
C}]_{\rm C}^M = \ \frac 32 \partial^M \left(\xi_{[1}^P \xi_2^Q \partial_P \xi_{3]Q} \right) + F^M (\xi_{[1}, \xi_2, \xi_{3]}) \ .\ee
The last term of this equations vanishes due to the constraint (\ref{FirstIntermediate}). Imposing this constraint and using (\ref{DefC}) the follow identity can be derived
\be 3[ \delta_{\xi_{[1}}, [ \delta_{\xi_{2}},\ \delta_{\xi_{3]}}   ] ]
 = \delta_{{\cal J}_{\rm C}(\xi_1,\xi_2,\xi_3)} \ .\label{preJacobi}\ee
Since the left hand side of (\ref{preJacobi}) vanishes because the
ordinary commutator satisfies the Jacobi identity, consistency requires that the Jacobiator generates trivial gauge transformations, and this leads to the second {\it closure} constraint
\be
H^M{}_N (\xi_{ 1},\xi_2, \xi_{3},V) =  \delta_{{\cal J}_{\rm C}(\xi_1,\xi_2,\xi_3)} V^M{}_N = \frac{3}{2}\partial^R \left(\xi_{[1}^P \xi_2^Q \partial_P \xi_{3]Q} \right) \partial_R V^M{}_N =  0 \ .\label{SecondIntermediate}
\ee
Again, for restricted configurations this is automatically satisfied, but it could admit other solutions.

The gauge transformations (\ref{gaugeK}) imply that $\cal R$ and $e^{-2d}$ transform as
\begin{equation}
\delta_\xi {\cal R} = \xi^M \partial_M {\cal R} + G(\xi,{\cal E}, d) \ , \ \ \ \ \ \
\delta_\xi e^{-2d} = \partial_M \left(\xi^M e^{-2d}\right)\ ,
\label{transf}
\end{equation}
with
\bea
G(\xi,{\cal E}, d) &=& - \partial^P \partial_M\xi_N \partial_P {\cal H}^{MN} - 2 \partial^P \xi_N \partial_P \partial_M {\cal H}^{MN}+ 4 \partial_P d \partial_M \partial^P\xi_N {\cal H}^{MN}\nn\\
&& + 4 \partial_P d \partial^P \xi_N \partial_M {\cal H}^{MN} + 4 \partial_N d \partial^P \xi_M \partial_P {\cal H}^{MN}\nn\\
&& + \frac 14 {\cal H}^{MN}\partial^P\xi_M \partial_P {\cal H}^{KL}\partial_N {\cal H}_{KL} - {\cal H}^{MN}\partial^P \xi_M \partial_P {\cal H}^{KL}\partial_K {\cal H}_{NL}\nn\\
&&   + 8 {\cal H}^{MN} \partial^P\xi_M \partial_P\partial_N d - 8 {\cal H}^{MN} \partial_M d \partial^P \xi_N \partial_P d \nn\\
&& -2 \partial_M\left(\partial^P \partial_P \xi_N {\cal H}^{MN}\right) + 4 \partial^P \partial_P \xi_ M\partial_N d {\cal H}^{MN} \nn\\
&& + \partial_P \xi^Q \partial_Q {\cal E}^a{}_M \partial^P {\cal E}^b{}_N S_{ab} \eta^{MN} + \partial_P \partial^N \xi^M {\cal E}^a{}_M \partial^P {\cal E}^b{}_N S_{ab} \nn\\
&& - \partial_P \partial^M \xi^N {\cal E}^a{}_M \partial^P{\cal E}^b{}_N S_{ab}
\ .\label{DeltaR}
\eea
Therefore, for $\cal R$ to transform as a scalar, the following {\it invariance} constraint must hold
\bea
\int d^K X\ e^{-2d}\ G(\xi,{\cal E}, d) = 0\ .\label{Invariance}
\eea

Also, for consistency, any subset of fields and gauge parameters allowed by the constraints must be such that under gauge transformations the transformed fields satisfy the constraints (\ref{FirstIntermediate}), (\ref{SecondIntermediate}) and (\ref{Invariance}) as well (i.e. the transformed fields must be also contained in the subset).

To summarize, consistency of gauge invariance in DFT requires two closure constraints (\ref{FirstIntermediate})  and (\ref{SecondIntermediate}) and the invariance constraint (\ref{Invariance}). All these restrictions involve contractions of two derivatives, and therefore vanish when the so-called weak and strong constraints hold, respectively
\be
 \partial^P \partial_P V^M{}_N = 0 \ , \ \ \ \ \  \partial^P W^R{}_S\ \partial_P V^M{}_N = 0 \ ,\label{strongconstraint}
\ee
where $W$ and $V$ denote any field or gauge parameter. Imposing these constraints was shown to be extremely restrictive. In particular, they only allow for restricted configurations of fields and gauge parameters that can be locally rotated to a frame in which they do not depend on the dual coordinates. Here we see that these constraints are sufficient to satisfy the closure and invariance conditions, but not necessary.

\section{Gauged Double Field Theory} \label{GDFT}

In this
section we compactify a $K$-dimensional DFT on a twisted torus of
dimension $d$. This leads to an effective theory defined on an
$N$-dimensional space with $N = K - d$, which we call GDFT. The resulting effective
action depends on a set of $N$ external coordinates denoted
$\mathbb{X}$, while the $d$ compact dimensions are referred to as
$\mathbb{Y}$. The compactification is defined by the (duality) twist matrix
\be
U^A{}_M (\mathbb{Y}) \in G \ ,  \ \ \ \ \label{twistMatrix}
\ee
that maps the $G$ indices $M,N$ of the parent DFT into the $G$ indices $A,B$ of the effective GDFT.
 One starts by proposing a reduction ansatz
that specifies the internal and external dependence of any arbitrary tensor
\be V^M{}_N(\mathbb{X},\mathbb{Y}) = (U^{-1})^M{}_A(\mathbb{Y})\
 \widehat{V}^A{}_B(\mathbb{X})\ U^B{}_N(\mathbb{Y}) \ ,\label{AnsatzTensor}\ee
where $V$ can denote either a field or a gauge parameter.
The notation is such that the hatted tensors $ \widehat{V}^A{}_B$ only depend on the
coordinates $\mathbb{X}$, and correspond respectively to dynamical objects or gauge parameters in the effective action.  The internal dependence is  fixed by the twist matrix (\ref{twistMatrix}), and has no dynamics.

It must then be demanded that the $\mathbb{Y}$-dependence factorizes out of the gauge transformations and the action. The information on the internal space does not disappear however, and is encrypted in gaugings, which are $\mathbb{Y}$-independent combinations of derivatives of the twists (\ref{twistMatrix}).

At this point there are two conditions that can be imposed on the ansatz (\ref{AnsatzTensor}). One is that Lorentz invariance of the $N$-dimensional effective
action is preserved, which requires that Lorentzian coordinates
remain untwisted, so any $\mathbb{X}$-dependent quantity
$\widehat g(\mathbb{X})$ must satisfy
\be (U^{-1})^M{}_A \partial_M \widehat g(\mathbb{X}) = \partial_A \widehat g(\mathbb{X})\ .
\label{restriction}\ee
Also, we restrict the set of external and internal coordinates in such a way that if a given coordinate is external (internal), its dual must be external (internal). This can be imposed through
\be
\partial^P U^{A}{}_M\ \partial_P \widehat g(\mathbb{X}) = 0 \ ,\label{restriction2}
\ee
and it ensures that the effective GDFT is also formally doubled.

\subsection{The effective action}
To obtain the effective action of GDFT, we start by twisting the generalized metric, and provide a scalar twist $\lambda(\mathbb{Y})$ to the $d$ field
\be {\cal E}^a{}_{M} (\mathbb{X},\mathbb{Y}) = U^B{}_M (\mathbb{Y})
\widehat{\cal E}^a{}_{B} (\mathbb{X} ) \ , \ \ \ \
\  \ d(\mathbb{X},\mathbb{Y}) = \widehat{d}(\mathbb{X}) + \lambda
(\mathbb{Y})\ .\label{Ansatz}\ee

When (\ref{Ansatz}) is inserted  in (\ref{action})
the action of GDFT is obtained
\bea S_{GDFT}= v \int d^{N}\mathbb{X}   \ e^{-2 \widehat d}\
\left({\cal R} ({ \widehat{\cal E} }, \widehat d ) + {\cal R}_f
(\widehat{\cal E}, \widehat d )\right)\ , \label{action2} \eea where $v$ is defined
by
\begin{equation}
 v = \int d^d \mathbb{Y} e^{-2 \lambda(\mathbb{Y})}\ , \label{Yfactor}
\end{equation}
and ${\cal R}_f$ is the gauged part of the action \bea {\cal R}_f &=& - \frac{1}{2}
f^A{}_{BC} \widehat{\cal H}^{BD} \widehat{\cal H}^{CE}
\partial_D \widehat{\cal H}_{AE} - \frac{1}{12} f^A{}_{BC}f^{D}{}_{EF}\widehat{\cal H}_{AD}\widehat{\cal H}^{BE}\widehat{\cal H}^{CF}
\nn\\ && - \frac{1}{4} f^A{}_{BC} f^B{}_{AD} \widehat{\cal H}^{CD} - 2 f_A \partial_B \widehat {\cal H}^{AB} + 4 f_A \widehat {\cal H}^{AB}\partial_B \widehat d
- f_A f_B \widehat {\cal H}^{AB}\ ,\label{DeformedRicci} \eea
where
\be
\widehat {\cal H}_{AB} (\mathbb{X})= \widehat {\cal E}^a{}_A(\mathbb{X})\ S_{ab}\ \widehat{\cal E}^b{}_B (\mathbb{X})\ .
\ee
All the dependence on the twists enters
in the effective action through the overall scale $v$ in (\ref{Yfactor}) and the gaugings
\bea f_{ABC} &=&  3
\eta_{D[A}(U^{-1})^M{}_{B}(U^{-1})^N{}_{C]}
\partial_M U^D{}_N \ ,\nn\\
f_A &=& \partial_M (U^{-1})^M{}_A - 2 (U^{-1})^M{}_A \partial_M \lambda \ ,\label{gaugings}\eea
which are taken to be constant. Note that
the restriction (\ref{restriction}) implies the following
consistency constraint
\be f^A{}_{BC}\ \partial_A\ \widehat g(\mathbb{X}) = 0 \ , \ \ \ \ \ f^A\ \partial_A\ \widehat g(\mathbb{X}) = 0 \ .\label{LorentzInv}\ee
This states that the
$N$-dimensional Lorentz invariance is not broken in the effective
action by the gaugings.

If we had not included the last term in the definition of the generalized Ricci scalar (\ref{R}), this action would involve an additional term proportional to the twists $U$, which are $\mathbb{Y}$-dependent. Since we have demanded all the $\mathbb{Y}$-dependence to factorize out, we are forced to include such additional term.

\subsection{Gauge symmetries and constraints} \label{sec:constraintsGDFT}

The gauge parameters $\widehat  \xi^A$ in GDFT are obtained by twisting the gauge parameter $\xi^M$ of the parent DFT
\be \xi^M (\mathbb{X},\mathbb{Y}) = (U^{-1})^M{}_A(\mathbb{Y})\
 \widehat{\xi}^A (\mathbb{X}) \ .\label{twistedparameters}\ee
The effective gauge transformations $\widehat{\delta}_{\widehat\xi}$ are defined through the relation
\be  {\delta}_{\xi} V^M{}_N = (U^{-1})^M{}_A\ U^B{}_N\ \widehat{\delta}_{\widehat \xi} \widehat{V}^A{}_B \ , \label{DefGauge}\ee
which guarantees that the transformed tensors maintain the ``Scherk-Schwarz'' structure (\ref{AnsatzTensor}).
After some algebra the following gauge transformation can be extracted from (\ref{DefGauge})
\bea \widehat{\delta}_{\widehat{\xi}} \widehat{V}^A{}_B &=&
\widehat{\xi}^C
\partial_C \widehat{V}^A{}_B + (\partial^A {\widehat{\xi}}_C - \partial_C {\widehat{\xi}}^A)
\widehat{V}^C{}_B +
(\partial_B \widehat{\xi}^C -\partial^C \widehat{\xi}_B) \widehat{V}^A{}_C \nn\\
&&- f^A{}_{CD}{\widehat{\xi}}^C \widehat{V}^D{}_B +
f^D{}_{CB}\widehat{\xi}^C {\widehat{V}}^A{}_D\ . \label{twistedGaugetransf}\eea
Note that the first line is the usual gauge transformation of the ungauged theory.
Again, for consistency the gaugings must be constant to ensure that gauge transformations do not reintroduce $\mathbb{Y}$-dependence.
For the fields of GDFT the gauge transformations (\ref{twistedGaugetransf}) read
\bea\widehat \delta_{\widehat\xi} \widehat{\cal E}^a{}_B &=& {\widehat\xi}^C
\partial_C \widehat{\cal E}^a{}_B + (\partial_B {\widehat\xi}^C -
\partial^C {\widehat\xi}_B)\widehat{\cal E}^a{}_C + f^D{}_{CB}{\widehat\xi}^C \widehat{\cal E}^a{}_D \ ,
\label{EffectiveGaugeInv}
 \eea
and
\bea \widehat\delta_{\widehat\xi} \widehat d(\mathbb{X}) &=&
{\widehat \xi}^A
\partial_A \widehat d - \frac 12 \partial_A \widehat \xi^A - \frac 12 f_A \widehat\xi^A \ ,\nn\\
\widehat\delta_{\widehat \xi} e^{-2\widehat d} &= & {\delta}_\xi
e^{-2d} =
\partial_A \left(\widehat \xi^A e^{-2\widehat d}\right)  + f_A \widehat \xi^A e^{-2\widehat d} \ .\label{GaugeInvDilaton}\eea

The C-bracket defines a twisted effective bracket, which we call
$f$-bracket $[ \ ,\ ]_f$, through the relation
\be [\xi_1  , \ \xi_2]^M_{\rm C} = (U^{-1})^M{}_A\ [\widehat \xi_1 ,\ \widehat
\xi_2]_f^A\ .\ee
The $f$-bracket preserves the structure of the C-Bracket but receives an extra contribution from the gaugings
\be [\widehat \xi_1,\ \widehat \xi_2]_f^A = [\widehat \xi_1,\ \widehat \xi_2]_{\rm
C}^A - f^A{}_{BD}\ \widehat \xi_1^B \widehat \xi_2^D \ .\label{fbracket}\ee
Closure of the effective gauge transformations also gives rise to the $f$-bracket plus twisted closure constraints (cf. equation (\ref{DefC}))
  \be \left[  \widehat{\delta}_{\widehat \xi_1}, \widehat {\delta}_{ \widehat\xi_2} \right] \widehat V^A{}_{ B} =
  \widehat{\delta}_{[ \widehat\xi_1, \widehat\xi_2]_f} \widehat V^A{}_{B} - \widehat F^A{}_B (\widehat \xi_1,\widehat \xi_2,\widehat V)\ . \ee
Using that the gaugings are constant, we get after some algebra the first twisted closure constraint
\bea
\widehat F^A{}_B (\widehat \xi_1,\widehat \xi_2,\widehat V)&=& U^A{}_M (U^{-1})^N{}_B F^M{}_N (\xi_1, \xi_2, V) \label{TwistedConstraints}
\\ &=& \widehat\xi^C_{[1} \partial^D \widehat\xi_{2]C} \partial_D \widehat  V^A{}_B + 2 \partial_D \widehat\xi^C_{[1} \partial^D \widehat\xi_{2]B}\widehat V^A{}_C+ 2 \partial_D \widehat\xi_{[1 C} \partial^D \widehat\xi_{2]}^A\widehat V^C{}_B\nn\\ && - 3 f_{F[CD} f^F{}_{E]}{}^A \widehat \xi^C_{[1}\widehat \xi^D_{2]}\widehat V^E{}_B- 3 f^{F[CD} f_F{}^{E]}{}_B \widehat \xi_{[1 C}\widehat \xi_{2]D}\widehat V^A{}_E \ =\ 0\ , \nn
\eea
which can be obtained either by twisting the first closure constraint in the higher dimensional DFT, or by demanding that two transformations (\ref{twistedGaugetransf}) reproduce a unique transformation. The second line in (\ref{TwistedConstraints}) is $F(\widehat \xi_1, \widehat\xi_2, \widehat V)$, and the last line is generated by gaugings that interestingly are arranged in a Jacobi-like form.

The Jacobiator $\widehat{\cal J}_f$ of the $f$-bracket is given again by a total derivative
\be\widehat {\cal J}_f (\widehat \xi_1,\widehat \xi_2,\widehat \xi_3)^D =  \frac 32 \partial^D \left(\widehat\xi_{[1}^A \widehat\xi_2^B \partial_A \widehat\xi_{3]B}  + \frac{1}{3} f_{ABC}
\widehat \xi_1^A \widehat \xi_2^B \widehat \xi_3^C \right) \ ,\ee
and demanding that it generates trivial gauge transformations leads to the second twisted closure constraint
\bea
\widehat H^E{}_F (\widehat \xi_{ 1},\widehat \xi_2,\widehat \xi_{3},\widehat V) &=& U^E{}_M (U^{-1})^N{}_F H^M{}_N (\xi_{ 1},\xi_2, \xi_{3}, V)  \nn\\
&=& \frac 32\partial^D \left(\widehat\xi_{[1}^A \widehat\xi_2^B \partial_A \widehat\xi_{3]B}  + \frac{1}{3} f_{ABC}
\widehat \xi_1^A \widehat \xi_2^B \widehat \xi_3^C \right) \partial_D \widehat V^E{}_F\ =\ 0 \ .\label{SecondTwistedClosure}
\eea
Again this expressions can be found either by twisting their analogs in the higher dimensional DFT, or by direct computations using the $f$-bracket and the  constraint (\ref{TwistedConstraints}).

The  deformed Ricci scalar (\ref{action2}) can be shown to transform as follows under gauge transformations (\ref{EffectiveGaugeInv}) and (\ref{GaugeInvDilaton})
\be
\widehat\delta_{\widehat \xi} \left({\cal R} ({ \widehat{\cal E} }, \widehat d ) + {\cal R}_f
(\widehat{\cal E}, \widehat d )\right)= \widehat \xi^A \partial_A \left({\cal R} ({ \widehat{\cal E} }, \widehat d ) + {\cal R}_f
(\widehat{\cal E}, \widehat d )\right) + G(\widehat\xi,\widehat{\cal E},\widehat d) + G_{f} (\widehat\xi,\widehat{\cal E},\widehat d)\ ,\label{DeformedRicciScalar}
\ee
where
\bea
G_f(\widehat\xi,\widehat{\cal E},\widehat d) &=& G(\xi,{\cal E}, d) -  G(\widehat\xi,\widehat{\cal E},\widehat d)  \nn\\&=&\frac{1}{2} \widehat{\cal H}^{AB} \partial^D \widehat \xi_A \partial_D \widehat {\cal H}^{EF} \widehat{\cal H}_{GF} f^G{}_{BE}+ \frac{1}{2} \widehat \xi_C \widehat {\cal H}^{AB}\widehat {\cal H}^{EF}\widehat {\cal H}^{GH} f_{D[A}{}^C f^D{}_{GE]} f_{BFH} \nn\\
&& - \frac{3}{2} \widehat \xi^C \widehat {\cal H}^{AB} \widehat {\cal H}^{EF} \partial_B\widehat {\cal H}_{GF} f_{D[AC} f^{D}{}_{E]}{}^G - \frac{3}{2} \widehat \xi^G \widehat {\cal H} ^{DH} f^{AC}{}_{[B} f_{GH] C} f^B{}_{AD}\nn\\ &&- f_A{}^{BC} \widehat{\cal E}^a{}_B \partial_D \widehat{\cal E}^{b}{}_C S_{ab} \partial^D\widehat\xi^A\ ,
\label{Gf}
\eea
provided the gaugings $f_A$ vanish. Let us briefly explain in the context of the higher dimensional theory why this should be the case. The parent DFT transforms as a total derivative involving the duality twists, which are not globally well defined. Therefore, to ensure gauge invariance, one should impose as a constraint for the duality twists that such derivative vanishes. Since it is proportional to $f_A$, these gaugings must be set to zero to ensure gauge invariance in the effective theory. In \cite{Geissbuhler:2011mx} it was shown how to introduce $f_A$ deformations through modifications of the ansatz that involve a warp factor. We will not consider this possibility here.
Due to (\ref{DeformedRicciScalar}) the action (\ref{action}) is then gauge invariant when the following twisted invariance constraints hold
\be
v \int d^N \mathbb{X} \ e^{-2 \widehat d} \ \left[G(\widehat\xi,\widehat{\cal E},\widehat d)  + G_f(\widehat\xi,\widehat{\cal E},\widehat d)\right] = 0 \ , \ \ \ \ f_A  =  0\ . \label{EffInvConstraints}
\ee

The action of GDFT formally preserves the $G$ covariance of the original theory. However, the gaugings explicitly break this group and  gauge a subgroup of it. It they were treated as spurions, and rotated under $G$, both the action and gauge transformations would transform covariantly. This is in complete analogy with gauged supergravities \cite{Samtleben:2008pe}. For a given gauging, the effective theory contains vectors preserving  a non-Abelian gauge symmetry, and this is the reason why we call this theory Gauged DFT. A concrete realization of this fact is given in Section \ref{Examples}, where we analyze the embedding of the non-Abelian heterotic supergravity in GDFT.

\subsection{Solutions to the constraints}\label{solutions}

In this section we present a possible family of solutions  to the effective constraints (\ref{TwistedConstraints}), (\ref{SecondTwistedClosure}) and (\ref{EffInvConstraints}). We require the effective tensors $\widehat V$ and $\widehat W$ to satisfy the weak and strong constraints on the external space
\be
\partial_E  \partial^E \widehat V^{A}{}_B  = 0\ , \ \ \ \ \ \ \partial_E \widehat V^{A}{}_B \partial^E \widehat W^{C}{}_D  = 0\label{ExternalStrong} \ .
\ee
In this case, (\ref{TwistedConstraints}) dictates that the gaugings should satisfy Jacobi identities
\be
f_{E[AB} f_{C]D}{}^E = 0 \label{JacobiId} \ .
\ee
The other constraints, equations (\ref{SecondTwistedClosure}) and  (\ref{EffInvConstraints}) are then automatically satisfied. We now argue that in general these configurations do not satisfy the strong and weak constraints (\ref{strongconstraint}).

It is convenient to define the tensor
\be
\Omega_{ABC} = \eta_{CD}(U^{-1})^{M}{}_{A}(U^{-1})^{N}{}_{B} \partial_M U^D{}_N = - \Omega_{ACB} \ ,
\ee
related to the gaugings through
\be f_{ABC} = 3\Omega_{[ABC]}\ .\ee
In terms of this object, the strong constraint for the duality twists
\be
\partial_P U^A{}_M \partial^P U^B{}_N = 0\ ,
\ee
implies
\be
\Omega_{EAB}\Omega^E{}_{CD} = 0 \ ,\label{strongrewritten}
\ee
while Jacobi identities read \cite{Geissbuhler:2011mx}
\be
f_{E[AB}f^E{}_{C]D} = \Omega_{E[AB}\Omega^E{}_{C]D}  - 4 U^E{}_M\eta_{E[A} \partial^M f_{BCD]} = 0 \ .\label{Jacobirewritten}
\ee
Since the gaugings are constant, the last term in (\ref{Jacobirewritten}) vanishes. Given that (\ref{strongrewritten}) appears antisymmetrized in (\ref{Jacobirewritten}), the strong constraint suffices to satisfy all the effective constraints, but it is not necessary.

This observation was also made in \cite{Geissbuhler:2011mx}, and the fact that the strong constraint is never used in the compactification procedure was previously noted in \cite{Aldazabal:2011nj}. In the light of our results, we now see that relaxing the strong constraint on the internal space is a perfectly consistent choice.

As for the weak constraint, we note that if we require the twist to satisfy it, given that $U\in G$, we get that the following condition holds
\be
\Omega^{CDA} \Omega_{CD}{}^B = 0 \ ,
\ee
which is not required by our constraint (\ref{JacobiId}). So similarly to the strong constraint, we get that the weak constraint is more restrictive than what we need for invariance of the action and closure of the algebra.

We would also like to point out the difficulty in relaxing the strong constraint while keeping the weak. In fact, for a set of fields and gauge parameters satisfying the weak constraint to be consistent, under gauge transformations the transformed fields must also satisfy this restriction. However, the transformed fields are given by products of (un-transformed) fields and gauge parameters, so for the weak constraint to annihilate the transformed field, the untransformed ones must obey strong-like constraints. Let us also comment on what would happen if the weak constraint were imposed on the solutions (\ref{ExternalStrong})-(\ref{JacobiId}). Notice that if the generalized metric together with the gauge parameters were restricted to satisfy the weak constraint, one obtains using (\ref{restriction2})
\bea
\partial_M \partial^M {\xi}_{P} &=& \partial_M \partial^M U^A{}_P\ \widehat{\xi}_{A} + U^A{}_P\ \partial_M \partial^M\widehat{\xi}_{A}  = 0\ ,\label{weakgauge}\\
\partial_M \partial^M {\cal H}_{PQ} &=& \partial_M \partial^M U^A{}_P\ \widehat{\cal H}_{AB}\ U^B{}_Q +  U^A{}_P\ \widehat{\cal H}_{AB}\ \partial_M \partial^M U^B{}_Q \nn\\ && +  2 \partial_M  U^A{}_P\ \widehat{\cal H}_{AB}\ \partial^M U^B{}_Q + U^A{}_P\ \partial_M \partial^M\widehat{\cal H}_{AB}\ U^B{}_Q  = 0 \ . \label{weakmetric}
\eea
The last terms in these equations vanish due to (\ref{ExternalStrong}). Also, since the effective gauge parameters are allowed to take values in any direction, then (\ref{weakgauge}) implies that the first two terms of (\ref{weakmetric}) vanish also. Therefore, the third term in (\ref{weakmetric}) must vanish
\be
\Omega_{EA(B} \Omega^{E}{}_{C)D} = 0\ .
\ee
Since $\Omega^{E}{}_{AB} = - \Omega^E{}_{BA}$, we conclude that $\Omega_{EAB}\Omega^E{}_{CD} = \Omega_{E[AB}\Omega^E{}_{CD]}$, which vanishes due to the Jacobi identities and constancy of the gaugings (\ref{Jacobirewritten}). Therefore, when the weak constraint is combined with the Jacobi identities, the strong constraint (\ref{strongrewritten}) must hold for the duality twists in the solutions (\ref{ExternalStrong})-(\ref{JacobiId}).

Let us stress that the weak constraint is derived in string theory from a worldsheet analysis in the context of Kaluza-Klein compactifications of closed strings on tori. It is possible, in light of our results based on a field theory analysis, that such constraint could be modified in Scherk-Schwarz compactifications of closed strings on more general spaces like the ones we consider in this paper. These issues deserve a better understanding and we hope to return to them in the future.

Although the weak and strong constraints can be relaxed on the internal space, consistency requires other constraints which can also be highly restrictive (although, in principle allow for truly double spaces). As we said, the weak and strong constraints must be replaced by the Jacobi identities (or more generally, the twisted constraints). The  condition that the gaugings must be constant restrict the possible configurations also. In addition, the duality twist is forced to be an element of $G$, and the condition $f_A=0$ obtained in  (\ref{EffInvConstraints}) (where $f_A$ are defined in (\ref{gaugings})) poses a condition on $\lambda(\mathbb{Y})$. Regarding global aspects, since the duality twist can be thought of as a generalized $K$-bein \cite{Dall'Agata:2007sr}, the differentials $U^A{}_M d \mathbb{Y}^M$ must be globally well defined up to identifications of coordinates. Finally, there are certain extra conditions on the algebra that have to be met such that one can consistently find a set of identifications to make the space compact (a discussion within solvable algebras can be found in \cite{Goi} and references therein). We will not discuss them here. Interesting works on how to generate gaugings from duality twists are \cite{Dall'Agata:2007sr, Andriot:2009fp}.

\subsection{Completing the action}

As pointed out  in \cite{Aldazabal:2011nj}, when the
effective action (\ref{action2}) is written in the form of gauged half-maximal supergravity \cite{Schon:2006kz}, it differs from the latter by the following term in the Lagrangian
\begin{equation}
-e^{-2 \widehat d}\frac{1}{6} f_{ABC} f^{ABC} \ .\label{ff}
\end{equation}
The absence of such term indicates that the theory actually  corresponds to a truncation of maximal supergravity \cite{Aldazabal:2011yz}.
Comparing our results with those of \cite{Hohm:2011ex}, we see that both actions differ by exactly the same term.

This term can have a higher dimensional origin in the following manifestly $G$-invariant additional piece in the action \cite{Geissbuhler:2011mx}
\begin{equation}
\triangle S_{DFT} = -\frac{1}{6} \int d^N \mathbb{X} d^d\mathbb{Y} e^{-2d}
{\cal F}_{abc} {\cal F}^{abc} = \int d^N \mathbb{X} d^d\mathbb{Y} e^{-2d} \triangle {\cal R}\ ,\label{Newterm}
\end{equation}
where the indices are contracted using the metric $\eta_{ab}$, which coincides with the metric of the group $G$. The tensor ${\cal F}_{abc}$ is given by
\begin{equation}
{\cal F}_{abc} = 3 S_{d[a}({\cal E}^{-1})^M{}_{b}({\cal
E}^{-1})^N{}_{c]}
\partial_M {\cal E}^d{}_N \ .\label{DynF}
\end{equation}

The consistency constraints (\ref{FirstIntermediate}), (\ref{SecondIntermediate}) and (\ref{EffInvConstraints}) do not require the term (\ref{Newterm}) to be zero.  The overall factor $-\tfrac{1}{6}$ in (\ref{Newterm}) is not fixed by the symmetries, but chosen by hand to reproduce the corresponding term in gauged half-maximal supergravities \cite{Aldazabal:2011nj, Geissbuhler:2011mx} and the heterotic string \cite{Hohm:2011ex}.

Under the gauge transformations (\ref{gaugeK}), this contribution to the generalized Ricci scalar transforms as
\be
\delta_\xi (\triangle {\cal R}) = \xi^M \partial_M (\triangle {\cal R}) + \triangle G (\xi, {\cal E})\ ,
\ee
where
\be
\triangle G (\xi, {\cal E}) = - \partial_P \xi^Q \partial_Q {\cal E}^a{}_M \partial^P {\cal E}^b{}_N \eta_{ab} \eta^{MN}  + 2 \partial_P\xi^N \partial^M {\cal E}^a{}_N \partial^P{\cal E}^b{}_M \eta_{ab}
\ ,
\ee
which contributes to the total $G$ in (\ref{DeltaR}).

When evaluated in the ansatz (\ref{Ansatz}),
the tensor (\ref{DynF}) splits into the contributions
\be
{\cal F}_{abc} = \widehat{\cal F}_{abc} + f_{DEF} \ (\widehat {\cal E}^{-1})^D{}_a (\widehat{\cal E}^{-1})^E{}_b (\widehat{\cal E}^{-1})^F{}_c\ ,
\ee
where \be \widehat {\cal F}_{abc} = 3 S_{d[a}(\widehat{\cal
E}^{-1})^E{}_{b}(\widehat {\cal E}^{-1})^F{}_{c]}
\partial_E \widehat{\cal E}^d{}_F\ ,\ee
and leads to two extra terms in the effective action
\bea \triangle S_{GDFT}& =& - \frac{1}{6} v \int d^N\mathbb{X}
e^{-2\widehat d}\left(f_{ABC} f^{ABC} + \widehat {\cal F}_{abc}
\widehat {\cal F}^{abc} \right)\nn\\&=& - \frac{1}{6} v \int d^N\mathbb{X}
e^{-2\widehat d}\left(\triangle R_f +  \triangle R(\widehat{\cal E})  \right)\ .\label{DeltaSeff}\eea
Under the effective gauge transformations (\ref{EffectiveGaugeInv}), the last term in (\ref{DeltaSeff}) transforms as
\bea
\widehat\delta_{\widehat \xi} \triangle R(\widehat{\cal E})  = \widehat\xi^C \partial_C \triangle R(\widehat{\cal E}) + \triangle G(\widehat \xi, \widehat{\cal E})+ \triangle G_f(\widehat \xi, \widehat{\cal E})
\eea
with
\be
\triangle G_f(\widehat \xi, \widehat{\cal E}) = \triangle G(\xi,{\cal E}) - \triangle G(\widehat \xi, \widehat{\cal E}) = f_A{}^{BC} \widehat {\cal E}^a{}_B \partial_D \widehat {\cal E}^b{}_C \partial^D \widehat \xi^A \eta_{ab}\ ,
\ee
contributing to $G_f$ in (\ref{Gf}). The first term in equation (\ref{DeltaSeff}) gives precisely the required contribution to complete the results of \cite{Hohm:2011ex}. It can be checked that the second term in (\ref{DeltaSeff}) vanishes using a standard parametrization (lower triangular for the case of $O(K/2, K/2)$) for the $K$-bein (see \cite{Dall'Agata:2007sr, Grana:2008yw}) if the weak and strong constraints are satisfied on the external space. In this situation, whenever $\widehat {\cal F}_{abc}$ in non-zero, $\widehat {\cal F}^{abc}$ vanishes. To understand this intuitively, one can think of an analogous situation for the first term. This terms corresponds to contractions between geometric and non-geometric fluxes, and is zero in situations where the weak and strong constraints apply to the internal space  and the twist is given a geometric parametrization. The setup in the next section is precisely the one under which the second term in (\ref{DeltaSeff}) vanishes.

\section{Heterotic String as a GDFT}\label{Examples}

In this section we would like to revisit the results of \cite{Hohm:2011ex}, and show how the non-Abelian heterotic strings \cite{Gross:1985rr} can be embedded in GDFT  (see \cite{Andriot:2011iw} for a discussion of the Abelian case). The first re-writing of the heterotic in a $G$ covariant way was done in \cite{Maharana:1992my}. It is not surprising that both the heterotic supergravity and half-maximal gauged supergravity in four dimensions can be embedded in GDFT, since the later can be obtained from dimensional reductions of the former (see \cite{Kaloper:1999yr} and references therein).

The novel feature here is that the structure constants of  the non-Abelian gauge group are not put in by hand, but have now a higher dimensional origin in the twist of a $d$-dimensional space. The starting  DFT has global symmetry group  $G = O(D,D+d)$, with $K = D + D + d$ and metric
\be\label{neweta} \eta_{MN} \ = \
\begin{pmatrix} 0 & 1_D & 0 \\
1_D & 0 & 0 \\ 0  &
0 &
1_d \end{pmatrix}\ , \ee
where $1_D$ is the $D \times D$ identity matrix.
The coordinates are fundamental vectors grouped according to
\be X^M = (\tilde x_i, x^i, y^\alpha)\ ,\ee
with $i=1,...,D$, $\alpha=1,...,d$. The generalized $(2D + d)\times (2D + d)$ metric is twisted as in (\ref{Ansatz}), where
the ``effective" generalized metric $\widehat {\cal H}$ is parameterized by the $D$-dimensional metric $g_{ij}$, the $D$-dimensional Kalb-Ramond form $B_{ij}$ and $d$ $D$-dimensional vectors $A_{i}{}^{\alpha}$
\be\label{newH}
\widehat {\cal H}_{MN}  =
   \begin{pmatrix} g^{ij} & - g^{ik} c_{kj}  & - g^{ik} A_{k \beta} \\
- g^{jk} c_{ki}  & g_{ij} + c_{ki} g^{kl} c_{lj} + A_{i}{}^{\gamma}
A_{j \gamma} & c_{ki} g^{kl} A_{l \beta}
 + A_{i \beta}  \\
 - g^{jk} A_{k \alpha} & c_{kj} g^{kl} A_{l \alpha} + A_{ j \alpha} & \delta_{\alpha \beta}
 + A_{k \alpha} g^{kl} A_{l \beta}
\end{pmatrix}\ ,\ee
with
 \be
  c_{ij} \ = \ B_{ij} + \frac{1}{2} A_{i}{}^{\alpha}
  A_{j}{}_{\alpha}\ .\label{c}
 \ee
The other generalized field  is defined by the  $G$ invariant
combination of the dilaton $\phi$ and the determinant of the metric $g$
\be e^{-2 \widehat d} = \sqrt{g} e^{-2 \phi}\ .\label{Dilaton} \ee

We consider the particular case in which $N = 2D = 10 + 10$, $d = 496$, and define $\mathbb{X} = (\tilde x, x)$ and $\mathbb{Y} = y$. The $d$ internal coordinates $y^\alpha$ are twisted, and to recover the low energy action of the heterotic string we will require no dependence on the dual external coordinates $\tilde x_i$. The  gaugings take the form
\be\label{fform}
   f^{A}{}_{BC} \ = \  \left\{
  \begin{array}{l l}
    f^{\alpha}{}_{\beta\gamma} & \quad \text{if\; $(A,B,C)=(\alpha,\beta,\gamma)$}\\
    0 & \quad \text{otherwise}\\
  \end{array} \right. \ ,
\ee
and are constrained by
\be
f^\delta{}_{[\alpha\beta} f^{\epsilon}{}_{\gamma]\delta} = 0\ .
\ee
In this case the constraints for external fields (\ref{TwistedConstraints}), (\ref{SecondTwistedClosure}) and (\ref{EffInvConstraints}) are trivially satisfied due to the fact
that the fields do not depend on the coordinates $\tilde x$. Notice that for the twisted internal space the metric (\ref{neweta}) is the identity.  Therefore, if the weak and strong constraints acted in the twisted internal space, they would remove all the dependence on $y^\alpha$, and therefore the gaugings would vanish, leaving the effective theory  with an Abelian gauge symmetry.

Under these assumptions and with these parameterizations,
the action of GDFT (\ref{action2}) completed with the extra contribution (\ref{DeltaSeff}) can be cast in the form
\be S_{\rm Het} =  v \int d^{10} x \sqrt{g} e^{-2 \phi}
\left[R + 4 \partial_i \phi \partial^i \phi - \frac{1}{12} \hat
H_{ijk} \hat H^{ijk} - \frac{1}{4} \delta_{\alpha \beta}
F^\alpha_{ij} F^{ij\beta}\right] \ ,\label{heterotic}\ee
where $R$ is the $10$-dimensional Ricci scalar, and we have defined
the curvatures
\be \hat H_{ijk} = 3 \left(\partial_{[i}B_{jk]} -
\delta_{\alpha\beta} A_{[i}{}^\alpha
\partial_j A_{k]}{}^\beta\right) + \delta_{\alpha \sigma}f^\sigma{}_{\beta\gamma} A_{[i}{}^\alpha A_j{}^\beta A_{k]}{}^\gamma\ , \ee
and
\be F^\alpha_{ij} = 2 \partial_{[i}A_{j]}{}^\alpha +
f^\alpha{}_{\beta \gamma} A_{[i}{}^\beta A_{j]}{}^\gamma \ .
\ee

Regarding the gauge symmetries of this theory, they are
inherited from the gauge symmetries of GDFT through
(\ref{EffectiveGaugeInv}). When evaluating the different components
of (\ref{EffectiveGaugeInv}) one finds the following transformation
properties for the fields
\bea \widehat \delta_{\widehat \xi} g^{ij} &=& \varepsilon^k \partial_k g^{ij} - 2
\partial_k \varepsilon^{(i} g^{j)k} \ ,\label{SimHetFrom}\\
\widehat \delta_{\widehat \xi} A_i{}^\alpha &=& \varepsilon^k \partial_k
A_i{}^\alpha +
\partial_i \varepsilon^k  A_k{}^\alpha +
\partial_i\Lambda^\alpha - f^\alpha{}_{\beta \gamma} \Lambda^\beta
A_i{}^\gamma \ ,\\ \widehat \delta_{\widehat \xi} B_{ij}&=& \varepsilon^k
\partial_k B_{ij} + 2 \partial_{[i} \varepsilon^k\ B_{j]k}  + 2  \partial_{[i} \tilde \varepsilon_{j]}
+  A_{[i}{}^\alpha \partial_{j]} \Lambda_\alpha \ ,\\
\widehat \delta_{\widehat \xi} \phi &=& \varepsilon^k \partial_k \phi \ , \label{SimHetTo}\eea
where we have parameterized $\widehat \xi^A = (\tilde \varepsilon
_i, \varepsilon^i, \Lambda^\alpha)$. This is precisely the way under
which the fields transform under diffeomorphisms parameterized by
$\varepsilon^i$, gauge transformations of the $B$-field parameterized by $\tilde
\varepsilon'_i = \tilde
\varepsilon_i - \tfrac 12 A_i{}^\alpha \Lambda_\alpha  $ and gauge transformations of the heterotic gauge fields parameterized by
$\Lambda^\alpha$ \cite{Hohm:2011ex}.

The action (\ref{heterotic}) together with its transformation laws
(\ref{SimHetFrom})-(\ref{SimHetTo}) are precisely those of the non-Abelian
 heterotic $10$-dimensional supergravity, provided  the
structure constants $\omega^a{}_{bc}$ of the $E_8 \times E_8$ or $SO(32)$
group of the Heterotic are identified with the
gaugings $f^\alpha{}_{\beta \gamma}$ in a basis in which the Killing
form $\kappa_{ab}$ coincides with the metric $\delta_{\alpha\beta}$. To make the connection explicit, we define a change of basis $\nu$ and the heterotic gauge coupling $g_0$
\begin{equation}
\kappa_{ab} = \nu^\alpha{}_{a} \delta_{\alpha \beta} \nu^\beta{}_{b}
\ , \ \ \ \   \delta_{\alpha\beta} = \nu^a{}_{\alpha} \kappa_{ab}
\nu^b{}_{\beta} \ , \ \ \ \    A_i{}^\alpha = \nu^\alpha{}_a A_i{}^a
\ , \ \ \ \   f_{\alpha\beta\gamma} = g_0 \nu^a{}_\alpha
\nu^b{}_\beta \nu^c{}_\gamma\omega_{abc}\ ,
\end{equation}
so that \be S_{\rm Het} =  v \int d^{10} x \sqrt{g} e^{-2
\phi} \left[R + 4 \partial_i \phi
\partial^i \phi - \frac{1}{12} \hat H_{ijk} \hat H^{ijk} -
\frac{1}{4} \kappa_{ab} F^a_{ij} F^{ij b}\right]\ , \ee
with
\begin{eqnarray} \hat H_{ijk} &=& 3 \left(\partial_{[i}B_{jk]} - \kappa_{ab}
A_{[i}{}^a
\partial_j A_{k]}{}^b\right) + g_0 \kappa_{ad}\omega^d{}_{bc} A_{[i}{}^a A_j{}^b A_{k]}{}^c \ , \\
 F^a_{ij} &=& 2 \partial_{[i}A_{j]}{}^a + g_0\omega^a{}_{bc} A_{[i}{}^b
A_{j]}{}^c\ .\end{eqnarray}

\section{Conclusions}\label{conclusions}

We revisited the gauge symmetries of DFT, and derived necessary and sufficient consistency conditions for closure of gauge transformations, (\ref{FirstIntermediate}), (\ref{SecondIntermediate}) and gauge invariance of the action (\ref{Invariance}). These conditions select sets of fields and gauge parameters for which DFT is consistent. The only previously known solutions to these conditions obeyed the so-called weak and strong constraints (\ref{strongconstraint}), that restrict the possible configurations to be T-dualizable to a frame in which there is no dependence on dual coordinates. Here, we propose a configuration (\ref{AnsatzTensor}) \`a la Scherk-Schwarz, that should satisfy (\ref{JacobiId}) and (\ref{ExternalStrong}) for consistency, but not necessarily the weak or strong constraint.

On these configurations, DFT is effectively described by a lower dimensional Gauged DFT, a DFT-like theory deformed by gaugings that preserve the global covariance of the parent theory. The action (\ref{action2}), global symmetries, gauge symmetries (\ref{EffectiveGaugeInv}), (\ref{GaugeInvDilaton}), bracket (\ref{fbracket}) and constraints (\ref{TwistedConstraints}), (\ref{SecondTwistedClosure}) and (\ref{EffInvConstraints}) are all derived from those of DFT (\ref{ODDninv}), (\ref{action}), (\ref{gaugeK}), (\ref{Cbracket}), (\ref{FirstIntermediate}), (\ref{SecondIntermediate}), (\ref{Invariance}).

Particular examples of GDFT are gauged supergravities and the non-Abelian heterotic supergravities. The embedding of the former into GDFT was discussed in \cite{Aldazabal:2011nj, Geissbuhler:2011mx}, and the later in \cite{Hohm:2011ex}. We reviewed the results of \cite{Hohm:2011ex} in the context of dimensional reductions, showing that the heterotic supergravities can be effectively obtained from Scherk-Schwarz compactifications of ungauged DFT.

There are a number of questions that follow from our analysis. In particular, it would be nice to find the classes of solutions to the constraints we found on the DFT, and determine if there is a maximal class. On the other hand, at the level of the reduced GDFT, it would be interesting to classify all the possible solutions to the constraints on the duality twists, and determine all the possible orbits of gaugings that can be turned on through this procedure. Generalizations of our procedure could allow for more general gaugings, like those obtained by turning on $f_A$. Another task would be to find an explicit example or existence proof of truly doubled backgrounds satisfying the constraints of Section \ref{solutions}.

Finally, relaxing the strong constraint forces more ambitious questions upon us. Given that a background for which there is no frame where the dependence on the dual coordinates disappears, one cannot stay within a limit where higher string excitation modes are heavier than the modes kept here. In that case one wonders whether a formulation of DFT without the strong constraint really corresponds to some limit of string theory\footnote{We thank D. Waldram and B. Zwiebach for sharing their viewpoints with us.}.

\bigskip

{\bf \large Acknowledgments} {We thank G. Aldazabal, D. Andriot, W. Baron, E. Bergshoeff, G. Dibitetto, D. Geissbuhler, A. Guarino, O. Hohm, K. Narain, C.
Nu\~nez,  D. Roest, A. Rosabal, D. Thompson, D. Waldram and B. Zwiebach for enlightening discussions and correspondence. This work was supported by the ERC Starting Independent Researcher Grant 259133 - ObservableString and by ECOS-Sud France binational collaboration project A08E06.}

\end{document}